\documentclass[10pt, conference, compsocconf]{IEEEtran}
\usepackage{subfigure}
\usepackage{flushend}

\usepackage{amssymb}
\usepackage{amsmath}
\usepackage{graphicx, color}
\usepackage[vlined]{algorithm2e}

\newcommand{\cancel}[1]{}

\newcommand{\ie}{\emph{i.e.}\xspace}
\newcommand{\eg}{\emph{e.g.}\xspace}
\DeclareMathOperator*{\argmax}{arg\,max}
\usepackage{url}
\newcommand{\baseline}{\textbf{baseline}\xspace}

\newcommand{\greedy}{\textbf{stream-greedy}\xspace}
\newcommand{\p}{$\mathcal{P}$\xspace}

\begin{document}
%
\title{(Re)partitioning for stream-enabled computation}


\author{\IEEEauthorblockN{Erwan {Le Merrer}, Yizhong Liang}
\IEEEauthorblockA{Technicolor\\
Rennes, France\\
surname.name@technicolor.com}
\and
\IEEEauthorblockN{Gilles Tr\'edan}
\IEEEauthorblockA{LAAS/CNRS\\
Toulouse, France\\
gtredan@laas.fr}
}


\maketitle

\begin{abstract}

  Partitioning an input graph over a set of workers is a complex
  operation. Objectives are twofold: split the work evenly, so that
  every worker gets an equal share, and minimize edge cut to achieve a
  good work locality (\ie workers can work independently).
  Partitioning a graph accessible from memory is a notorious
  NP-complete problem. Motivated by the regain of interest for the
  stream processing paradigm (where nodes and edges arrive as a flow
  to the datacenter), we propose in this paper a stream-enabled graph
  partitioning system that constantly seeks an optimum between those
  two objectives. We first expose the hardness of partitioning using
  classic and static methods; we then exhibit the cut versus load
  balancing tradeoff, from an application point of view.

  With this tradeoff in mind, our approach translates the online
  partitioning problem into a standard optimization problem. A greedy
  algorithm handles the stream of incoming graph updates while
  optimizations are triggered on demand to improve upon the greedy
  decisions. Using simulations, we show that this approach is very
  efficient, turning a basic optimization strategy such as hill
  climbing into an online partitioning solution that compares favorably to 
  literature's recent stream partitioning solutions.

\end{abstract}


%
\IEEEpeerreviewmaketitle

\section{Introduction}
\label{introduction}

Contemporary big-data applications ingest considerable amounts of data
to produce meaningful and valuable information. 
As datasets keep on growing to unprecedented sizes, applications must
rely on efficient and scalable computation means. Graph-based
applications as social networks~\cite{engines}, search engines or
recommender systems~\cite{perso-pagerank} have to deal with giant and
constantly evolving networks of user or item interactions. As those
terabytes of data cannot be efficiently processed and served by a single machine
in the horizontal scalability model using commodity hardware, the
solution is to \textit{partition the interaction graph} onto multiples
machines for parallel computation and request handling~\cite{kdd12}.
As computation is fast with this scheme,
the dataset evolution could be incorporated seamlessly by the
application, so that fresh results are always available.


The \textit{MapReduce framework} allows to process massive amount of
information, in an offline manner~\cite{mr}. Since very recently,
there is a resurgence of interest about \textit{stream processing},
with the proposal of open platforms such as Storm~\cite{storm} or commercial ones like Google MillWheel~\cite{milky} or Amazon Kinesis~\cite{kinesis}. In this
framework, data is treated as a flow, and each flow element is
processed on the fly (and then possibly discarded). While more
restrictive that MapReduce, this allows for online computation.

As the raison d'\^etre of stream processing is to exhibit a low latency
in its operation, relying on offline partitioning methods is not an
option. As the graph structure continuously changes due to node/edge
creations, calling a procedure that recomputes a partitioning from
scratch at each change is overkill (see e.g. traditional approaches
as~\cite{balanced}). In other words, incremental approaches to
partitioning are mandatory.  In this light, datacenter applications
like Pregel~\cite{pregel} partition nodes onto machines based solely on
their IDs; this is apparented to dispatch those nodes at random. A
recent work proposes to load the graph stored on a disk as a stream of
nodes, and to use cheap heuristics for node placement over one of the
$k$ processing machines on the fly~\cite{kdd12}. Although the fact that this
approach handles nodes as they are read, it assumes a full knowledge
model, where the whole graph is accessible at a given time as the
input.




In this paper, we propose a system that receives incoming edges, and
places their endpoint nodes greedily in partitions, then performing
online partitioning.
Our system operates over a continuous flow of events
arriving at a datacenter, then suiting the stream processing
paradigm. Greedy placement is complemented with periodic partitioning
reconfigurations at runtime, using solely little feedback from the
application.



The contributions of this paper are:
\begin{itemize}
\item to exhibit \textit{(i)} the instability of optimal partitions
  created by static partitioning algorithms, if they are run each time
  few new edges are added to the current graph, and \textit{(ii)} the
  existence of a graph-related tradeoff between a well balanced graph
  (work is evenly divided among the parallel instances) and a low edge
  cut (workers should be able to process most of the requests
  using information local to their partition). Intuitively, this tradeoff
  pops up every time the system has to decide between favoring
  edge-cut at the price of well balancedness, or vice versa.
\item based on these observations, to consider the problem of
  stream-enabled graph partitioning as a standard mathematical
  optimization problem. In this problem, the optimization parameters
  are the well balancedness and the edge-cut, and the metric to optimize
  is the application performance, measured for instance by the average
  request processing time.
\item to propose a stream-enabled graph partitioner built upon these
  observations. The realistic simulations we conducted show that a
  standard greedy optimization is efficient compared to current state-of-the-art
  stream-enabled partitioner~\cite{kdd12}, while executing in a more
  restricted model. Thanks to the greedy nature of the
  optimization, this performance is achieved for cheap.
\end{itemize}


The remaining of this paper is structured as follows:
Section~\ref{model} exhibits the danger of seeking
optimal partitionings in the context of streamed graphs, before
presenting the model of execution considered. We then observe in
Section~\ref{tradeoff} that traditional graph partitioning metrics are
bound by a graph-dependent tradeoff that impacts application
performance. Based on these observations, Section~\ref{stream} models
the streamed-graph partitioning problem as an optimization problem,
and proposes a greedy online partitioning mechanism, along with
simulation results. A reconfiguration technique to be used
periodically at runtime is then presented. Section~\ref{illustration}
illustrates this partitioning scheme in an application context:
graph-based recommendation. We finally present Related Work in
Section~\ref{related} before we conclude.

\section{Model: processing over streamed graphs}
\label{model}

\subsection{Instability of Optimal Partitionings}
\label{sec:unst-optim-part}

This sections illustrates the difficulty of simply transposing traditional graph
partitioning metrics in the context of graph streaming. The first major problem
comes from the complexity of finding an optimal partitioning, which is
NP-complete~\cite{NP}. Considering the update rate of a streamed graph, such
problem would have to be solved for every increment, which is not realistic.

\begin{figure}[h]
  \centering
  \includegraphics[width=3cm]{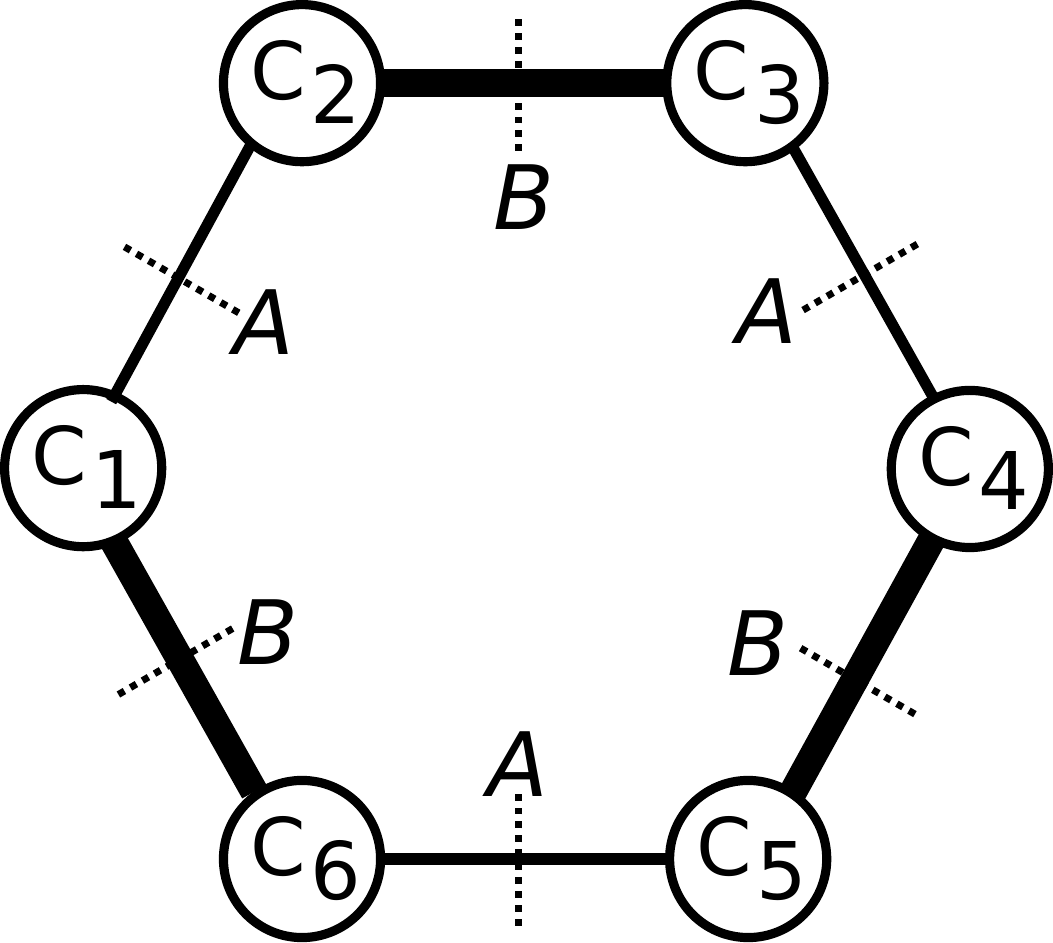}
  \caption{A ring graph for $k=3$. Depending on the order of new edge
    arrivals, this topology triggers unstable decisions by
    partitioning methods.}
  \label{fig:dynavice}
\end{figure}

Second problem stems from the unstability of such optimal partitioning. To
illustrate this, consider the following example. Assume that $k$ servers operate
on a graph composed of a ring of $2k$ fully-connected clusters $C_1,\ldots
C_{2k}$, with $ \vert C_i\vert = \frac{n}{2k}$. Assume for $i\in [1,k]$, we have $A$
links between $C_{2i-1}$ and $C_{2i}$, and $B$ links between $C_{2i}$ and
$C_{2i+1 [2k]}$, with $A,B < \frac{n}{2k}$. Figure~\ref{fig:dynavice} illustrates such topology in the case
$k=3$.

Since any graph partition cutting through a cluster $C_i$ would cut at least
$\frac{n}{2k}$ links, balanced cuts are either along $A$, either along
$B$. Assume $A=B-1$: the optimal assignment is to map $(C_{2i} \cup C_{2i+1[2k]})$
to server $i$. Now assume that two new edges arrive on the $A$ cut, for instance
between $C_1$ and $C_2$. The new optimal assignment is to map $(C_{2i} \cup
C_{2i-1[2k]})$ to server $i$. To reach this new optimal assignment,
$\frac{n}{2}$ nodes need to be transferred during the reconfiguration. Observe
that the arrival of two new edges along a $B$ cut (or the removal of two edges
along an $A$ cut) can now generate the same amount of reconfigurations.

Let us imagine a worst case scenario where originally $A=1$ and $B=2$,
and a stream of edges repeating the aforementioned scheme
($A,A,B,B,A,A,\ldots$) until $A=B=\frac{n}{2k}$.  The described
reconfiguration then happens $k\frac{n}{4k}$ times. This implies
$\frac{n}{2}\frac{n}{4k} = \Omega(n^2)$ node transfers. Although this
is a pathological worst-case scenario, bad situations cannot be
discarded in dealing with real world graphs either. This illustrates the need
for a specific approach to support graph updates, then adapted to the
streamed graph model we now define.

\subsection{Partitioning Model and Metrics for Streamed Graphs}
\label{sec:part-model-metr}

We consider a streamed graph model, where edges arrive continuously to
a central machine called the \textbf{partitioner}, \p. \p is in charge
of partitioning the streamed graph $G=(V,E)$ over a fixed set of $k$
machines or cores, according to the computing hardware
setup. Practically, it positions an incoming edge endpoints (\ie nodes)
on one or two of the machines that are hosting a partition of $G$,
under the form of adjacency lists. $G_{\infty}$ can be seen as the graph
resulting from the aggregation of all arrivals, at the end of times.
$P(i)$ denotes partition number $i$, and $|P(i)|$ the number of nodes
currently in $P(i)$.  Each partition has the capacity to host $C$
nodes. $\bigcup\limits_{i=1}^k P(i)$ then contains all nodes and edges
from $G$ seen so far.

We do not make any assumption on the order of arrival of elements in set $E$. The
system operates on edges, implying that if one endpoint (\ie one node)
is unknown, its creation is handled by the system. $\Gamma(v)$ denotes
the set of neighbors of node $v$.

\p maintains a table mapping all nodes seen so far (\ie edge
endpoints) to their current partition assignment (an integer $[1,k]$), for
being able to take greedy decisions on placement of incoming
edges. State maintained at \p is thus $O(|V|)$.


In this paper, we are interested in optimizing the average request processing
time of the application on top of which our partitioning system is
deployed. However, throughout the paper, we refer to two traditional
metrics of graph partitioning:
\begin{itemize}
\item \emph{Load balancing} is computed as the spread between the less and the
  more loaded of the $k$ partitions. It can be written as $$\frac{\min_{i \in [k]}(\vert
    P(i) \vert)}{\max_{i \in [k]}(\vert P(i) \vert)},$$ where 0 denotes a very uneven load
  between the partitions, and 1 denotes a perfect balancing.
\item \emph{Cut} is the fraction of edges that have endpoints located
  in different partitions. This represents an important quantity, as
  traversing such edge will require data to be exchanged among the
  machines, therefore adding latency to the request processing
  time. Formally, the cut is defined as: $$1-\frac{\vert \{(a,b) \in E
    \vert a\in P(i), b\in P(j), i\neq j \}\vert}{|E|}.$$ Again, 0
  denotes a non-desirable situation where all edges have endpoints in different
  partitions, and 1 denotes a partitioning cutting no edge.
\end{itemize}

\section{Partitioning: cuts, load and applications}
\label{tradeoff}

Let us first consider a coarse model of the environment of an
application 
running in a centralized setting. Upon arrival of a request $r$, this application
will consume two quantities: memory and CPU time. Let $m(r)$ and $c(r)$ be these
quantities. In this abstract model, we consider that the system knows instantly
at the arrival of $r$ what will be $m(r)$ and $c(r)$.

The system is able to provide memory and processing power at rate $\mu$ and
$\chi$ per time unit, respectively. Therefore, if $r$ is the only request on the
system, we consider that its processing will take the time required to gather
the required resources $t_r = m(r)/\mu +c(r)/\chi$.  As a simple model for
congestion, assume that $n_r$ requests are processed on the system at each
time. Then $t_r$ becomes $ n_r.(m(r)/\mu +c(r)/\chi)$: the system evenly splits
its processor and memory supply to all the requests, and side effects (such as
context switch) are neglected.

Now let us consider the same application running in a distributed
setting. The input of this application is a graph $G$: the memory
requirements of a request $r$ can now be expressed as a subgraph of $G$:
$m(r)\subset G$. We consider the following distributed setting: $k$ machines are
fully connected through synchronous equal links. Each machine $i$ has enough
memory to hold a subgraph $P(i)\subsetneq G$ such that $\{ P(i)\}_{1\leq i\leq k}$
forms a partition of $G$. 

\subsection{Analyzing a Simple Locality Model}
\label{sec:memory-model}

Let us assume the memory needs of a request consist in the $\ell$-hop
neighborhood of a node: $m(r) \simeq B(v,\ell)$, where $v$ is the center of
request $r$. We define $\ell$ as a measure of requests' \emph{locality}. Let us
illustrate this concept:
\begin{itemize}
\item the request ``get $v$'s neighbors'' has a locality of $0$: the neighbors of
  $v$ are known locally by $v$.
\item the request ``get $v$'s eccentricity'' has a locality of $D$, the graph
  diameter, as the most eccentric nodes are $D$ hops apart.
\item a damping random walk (jump with a probability $\alpha <1$) can be
  modeled by an ``expected'' locality (e.g. $\ell= \lceil -1/\log(\alpha) \rceil$).
\end{itemize}

Let $p$ be the machine holding node $v$, center of a request $r$ of locality
$\ell$. If $B(v,\ell) \subset P(p)$, all the required information to process $r$
is already available on $p$, the request processing time only depends on the
processing resources available on $p$. However, if $\exists q, B(v,\ell) \cap
P(q) \neq \varnothing$, then information will have to be fetched from machine
$q$, and the duration of this fetch will add up to the request processing time.

More formally, let $\lambda$ be the network latency induced by such fetching
operation. We consider that remote fetches cannot be made parallely, mostly
because in the streaming context the $\ell$ hop neighbors (and therefore the
partitions holding them) are not known in advance beside direct
neighbors. Therefore we model the processing time of request $r$ when processed
by $p$ as: $$t_r= \underbrace{\frac{c(r) \vert
    P(p)\vert}{\chi_p}}_{\text{computing time}}+ \underbrace{\lambda \vert \{ j \neq
p, \text{s. t. } P(j) \cap B(r,\ell) \neq \varnothing\} \vert}_{\text{information
gathering time}}.$$

Observe that the computing time contribution depends on the size of $P(p)$ since
the bigger the partition is, the more requests machine $p$ will have to serve in
parallel. The information gathering time also depends on $P(p)$ since the bigger
$P(p)$ is, the higher the chances are that $B(v,\ell) \subset P(p)$, therefore
reducing the information gathering time to 0.

Thus, we have here a first visible \textbf{tradeoff} the partitioning strategy has
to solve in order to minimize request compute time:
\begin{itemize}
\item Computations over small partitions are processed faster, since the load on the machine
  holding the partition is low, at the cost of higher information gathering
  costs.
\item Computation over big partitions are slower, but require on average less
  information fetching.
\end{itemize}
Now, considering that we have a fixed number of machines $k$, this tradeoff
translates in: \emph{shall we prefer to minimize the cut or to optimize the load
balancing ?}

\subsection{Graph-Related Tradeoff}
\label{sec:graph-relat-trad}

\begin{figure}[t!]
  \centering
  \includegraphics[width=3cm]{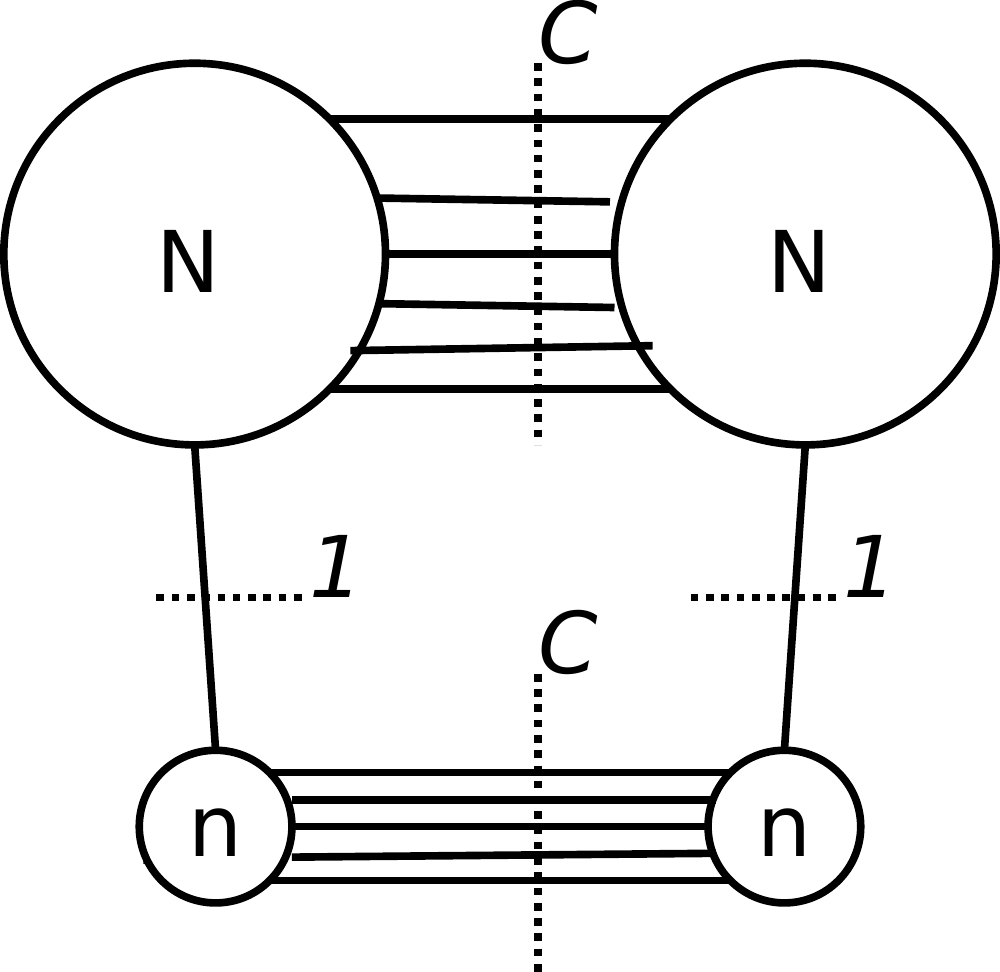}
  \caption{A vicious graph for partitioning methods. Two contradictory
    decisions can be made: favoring load balancing OR cut ratio.}
  \label{fig:vicous}
\end{figure}

With the aforementioned tradeoff in mind, consider the graph depicted
figure~\ref{fig:vicous}. This graph consists in $4$ fully connected clusters of
sizes $N,N,n$ and $n$. Clusters of equal size are connected by $C$ links, and
two links connect one cluster of size $N$ with one of size $n$. Assume that
$N>n$ and $n>C>1$. Two key observations are:
\begin{itemize}
\item Any exactly balanced bisection (\emph{i.e.} two partitions $G_1,G_2$ such
  that $|G_1| = |G_2| = (N+n)$) of the graph cuts at least $2C$ links. Let
  $P_{WB}$ such partition, symbolized by $C$s on Figure~\ref{fig:vicous}.
\item The graph is $2$-connex. The minimal edge-cut is $2$ and has a
  balancedness $\min (|G_1|,|G_2|)/\max (|G_1|,|G_2|)$ of $n/N$. Let
  $P_{MC}$ such partition, symbolized by $1$s on Figure~\ref{fig:vicous}.
\end{itemize}

Now let us compute the average request processing time $\mathbb{E}(t_r)$
centered on a node $v$. Since we have only two clusters, assuming
$\ell\in\{0,1\}$, computing the information fetch cost is easy. Let
$\mathcal{B}$ be the boundary of each cluster, and $\phi = c(r)$.

Assume well balancedness is preferred:
  \begin{align}
    \mathbb{E}(t_r | P_{WB}) &= \frac{\phi(n+N)}{\chi}+ \lambda \ell \Pr(v\in \mathcal{B}) \\
    &= \frac{\phi(n+N)}{\chi}+  \lambda \ell \frac{2C}{n+N}.\label{eq:1}
  \end{align}

Now assume cut minimization strategy is preferred:
  \begin{align}
    \mathbb{E}(t_r | P_{MC}) &= \Pr(v\in P_1)\frac{\phi|P_1|}{\chi}+
    \Pr(v\in  P_2)\frac{\phi|P_2|}{\chi} +\\& \lambda \ell \Pr(v\in \mathcal{B})\\
    &=  \frac{N}{n+N}\frac{2N\phi}{\chi}+ \frac{n}{n+N}\frac{2n\phi}{\chi} +
    \lambda \ell \Pr(v\in \mathcal{B}) \\
    &= \frac{2\phi (n^2+N^2)}{\chi(n+N)} + \lambda \ell \frac{2}{n+N}. \label{eq:2}
  \end{align}

If we compare those two quantities we have:
\begin{align}
  \label{eq:3}
    \mathbb{E}(t_r | P_{WB}) &\leq \mathbb{E}(t_r | P_{MC}) &&\Leftrightarrow \\
   \phi(n+N)^2 +2\ell \lambda  \chi C &\leq  2\phi(n^2+N^2) + 2\ell \lambda  \chi  &&\Leftrightarrow \\
    2\ell \lambda  \chi (C-1) &\leq \phi (n^2+N^2-2nN)  &&\Leftrightarrow \\
    2\ell \lambda  \chi (C-1) &\leq \phi (n-N)^2 \hfill &&\Leftrightarrow \\
        \frac{2\ell \lambda  \chi}{\phi} &\leq  \frac{(n-N)^2}{C-1}.
\end{align}
Therefore, in such a setting, one can draw the following conclusions:
\begin{itemize}
\item if the problem is only local ($\ell=0$), well balancedness is always faster.
\item under this last form, the inequality's left hand side only contains
  application and hardware dependent variables, that are unlikely to change
  along the progress of the stream. The inequality's right hand side only contains graph
  dependent variables: these are likely to evolve along the streaming progress.
\end{itemize}

This simple model illustrates a major problem of classical
graph-partitioning approaches when dealing with the incremental nature
of streamed graphs: the fastest partitioning strategy depends
\textbf{both} on the target \textbf{application} and on the target
\textbf{graph}. As at least the graph is unknown, meaning that we
cannot make strong assumptions on the evolution of its characteristics
or on the order on which will be received particular updates, we
therefore argue that an online exploration of this well
balancedness/min-cut tradeoff is mandatory for top-application performance.

\begin{figure}
\center
\includegraphics[width=.95\linewidth]{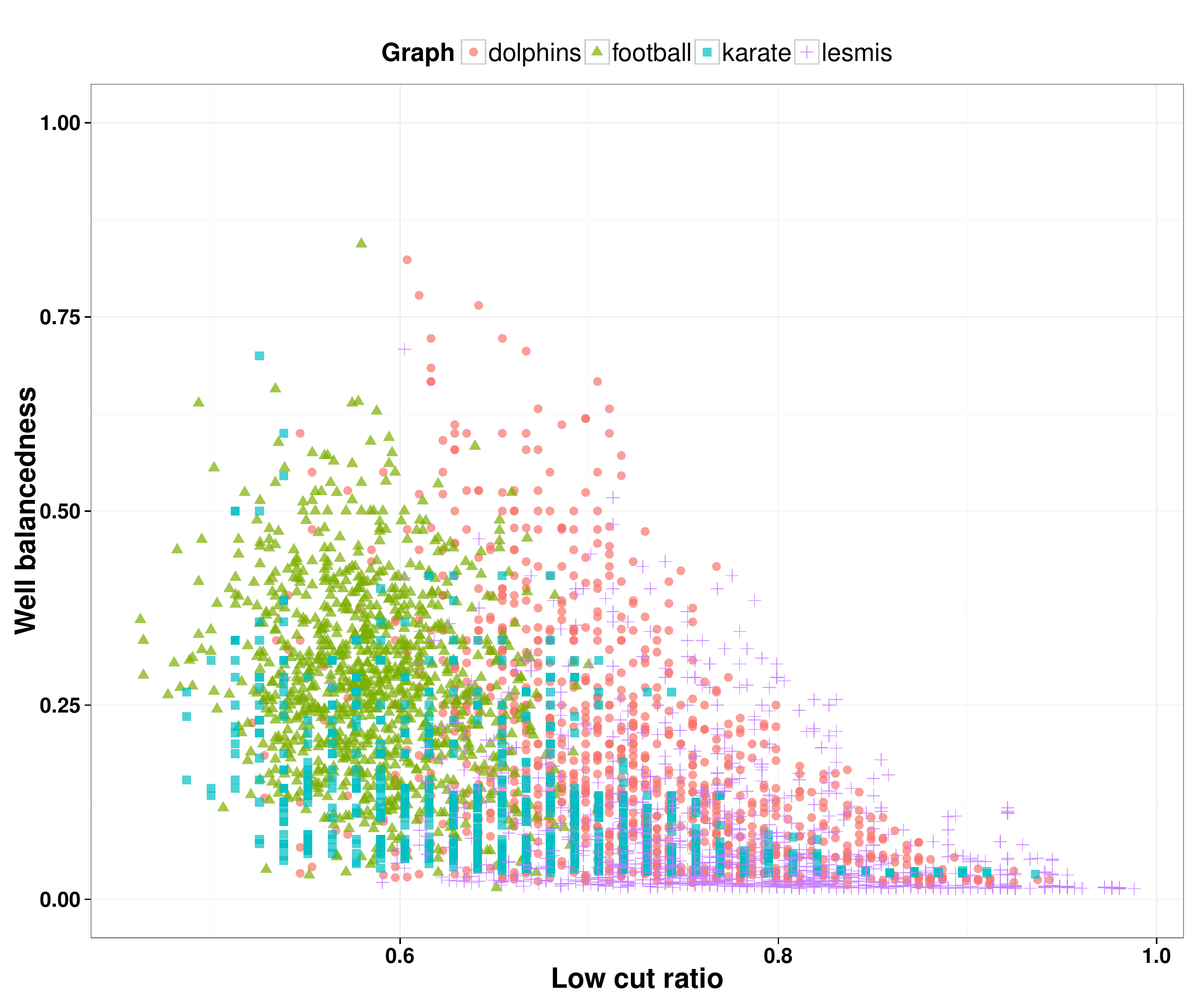}
\caption{Reachable configurations while partitioning $4$ different
  graphs, under the load balancing vs cut ratio tradeoff. Each point
  represents a particular configuration: each graph as it own
  particular set of ``good'' configurations (positions on the
  top-right envelope are the desirable ones).}
\label{fig:partitionings}
\end{figure}

 One might wonder, due to the artificial nature of the
  constructions used to illustrate this tradeoff, whether such
  situation do happen in practice. To answer this question, we took a
  set of standard small real-world topologies\footnote{Topologies are
    available at
    \url{http://www-personal.umich.edu/~mejn/netdata/}. The authors
    would like to thank M.E.J. Newman for providing these
    topologies.}, and randomly partitioned them into $4$ pieces, and
  measured the obtained well balancedness and cut size. We repeat the
  process $1,000$ times for each topology, therefore ``sampling'' the
  partitioning configuration
  possibilities. Figure~\ref{fig:partitionings} represents the
  obtained results. Each point represents a random partitioning, and
  although chances are low that optimal partitionings are represented
  on this figure, the point cloud associated with each topology
  represents the likely outcomes of partitionings. As we can see, they
  differ from one topology to the other: each topology allows its own
  tradeoff between well balancedness and low edge cut. 

\section{(Re)partitioning from streamed graphs}
\label{stream}

We have seen in previous sections that the hardness of partitioning
precludes a new partitioning iteration on the whole graph at each new
edge arrival. We have established that the partitioning achieving the
lowest request processing time has to find an optimum between a good
load balancing, and a low edge cut. Moreover, we have seen that such
optimum not only depends on the application, but also on the
characteristics of the streamed graph.

We now propose a greedy solution for partitioning and reconfiguration,
that handles increments as graphs are streamed.

\subsection{Global Framework Overview}
\label{sec:over-syst-overv}

\begin{figure}[h]
  \centering
  \includegraphics[width=.9\linewidth]{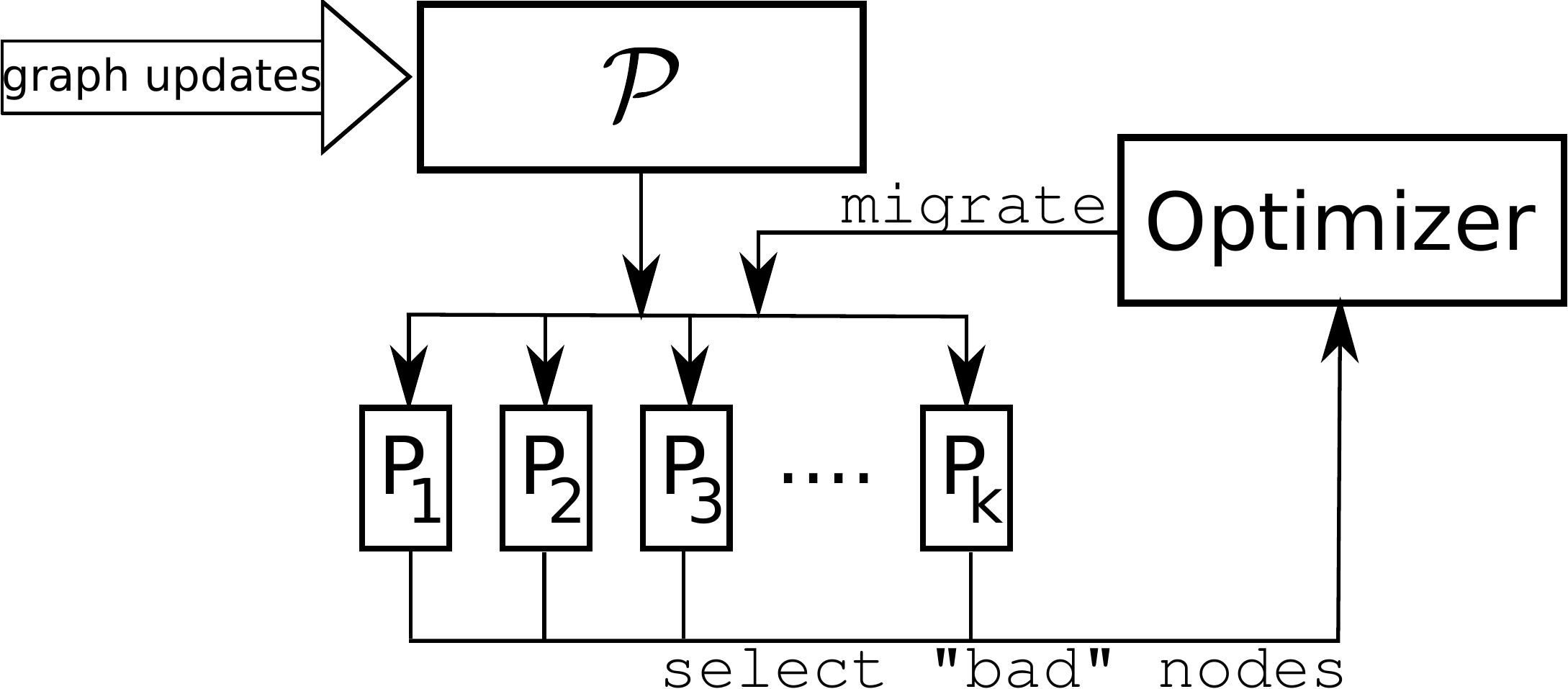}
  \caption{Overall system overview, with three logical components: the partitioner, the optimizer and the partition hosts.}
  \label{fig:overview}
\end{figure}

The framework we propose is the following one, as shown on
Figure~\ref{fig:overview}. The stream of incoming graph updates (edge
additions) is dispatched to the partitions by \p. Since the optimal
balance between load balanced partitions and low cut ratio depends on the
graph (which is constantly evolving) and the application (whose sweet
spot for performance is not necessarily known), and since the
continuous stream of updates has to be handled as quickly a possible, the
partitioner decides which partition to assign new edges to in a greedy
fashion. A logical optimizer monitors the state of the machines,
and periodically optimizes this greedy layout by choosing nodes in
partitions, and by migrating them. The choice of which nodes to move,
and of where to migrate them is driven by an optimization procedure
following two strategies: it tries to improve either the edge cut,
either the load balancing. The precise role of each component is
described afterwards.

\subsection{Stream-greedy: A Simple Greedy Partitioner}

We first introduce the two related work competitors for
efficient online partitioning.
\subsubsection{Related Approaches}
The common approach, due to its simplicity, is to partition at random.
Once a node arrives at the datacenter, a partition is selected
according to a modulo operation over the node identifier (itself being
pseudo-random), see \eg Pregel~\cite{pregel}. This often called
\textit{hashing}.

More advanced heuristics for partitioning a graph that is read from
disk as a stream of nodes, are presented in~\cite{kdd12}. They all
outperform hashing. The best performing heuristic is called
\textit{weighted deterministic greedy}, and is node driven. It consists of placing an
incoming node $v$ to the partition $j$ where it has the most edges,
weighting this by a linear penalty function based on the capacity of
the partition:

\begin{equation}
j = \argmax_{i \in [k]} (|P(i) \cap \Gamma(v)| (1 - \frac{|P(i)|}{C})).
\label{eq:baseline}
\end{equation}

All methods stream nodes in random order or in a breadth/depth first
search fashion. The major assumption of this technique is that each
streamed node comes with its complete (non-yet seen)
neighbor-list. This make this study not applicable to the stream
processing model where edges are received as events at a
datacenter along the application life (as opposed to nodes read from a
disk containing the whole graph).

We pick this deterministic greedy heuristic as the \baseline technique
to compare to in the rest of this paper.


\subsubsection{stream-greedy partitioner}

\begin{algorithm}[t!]
    Initialize $k$ partitions with capacity $C$\;
    For each incoming edge $e_{ij}$:\\
    \uIf{$\exists i~and~\exists j$}{
      $addLink(i,j)$
    }
    \uElseIf{$\exists i~and~\nexists j$}{
      Place $j$ in $P(i)$ if not full.
      Otherwise, place $j$ at the least occupied partition. $addLink(i,j)$
   }
    \Else{//i.e. $\nexists i~and~\nexists j$\\
      Place both $i$ and $j$ at the least occupied partition. $addLink(i,j)$
    }
\vspace{0.2cm}
\caption{\footnotesize \greedy heuristic for streamed graph partitioning.}
\label{greedy-alg}
\end{algorithm}


As we are bound to a restrictive model where edges are
received arbitrarily following the application logic, and because a
crucial point for system scalability is to propose a fast and
lightweight partitioning method at \p, we detail a simple and
intuitive greedy partitioner: when $e_{ij}$ arrives at the datacenter,
placement decision is made following the pseudo-code provided
on Algorithm~\ref{greedy-alg}.  $addLink(i,j)$ is a function triggered by
\p, that informs machines hosting $i$ and $j$ (they could be one
single machine) to link those two nodes via an edge in their
respective adjacency lists.

Computationally, this heuristic simply requires \p to perform
membership tests and cardinality operations over the mapping table
(for finding the least represented partition).



\subsection{Improving Current Partitioning}

To improve the current partitioning, the optimizer relies on two heuristics. The
first targets an improvement on the load balancing criterion, while the second
targets an improvement on the cut criterion. The simulation results presented
Section~\ref{sec:simulations} show that both heuristics are efficient most of the time: each one
improves one metric without significant degradation of the other.

Both heuristics select nodes from machines based on their ``badness'',
which measures each node's individual contribution to the size of the
cut. Rationale behind using the same selection being that even if load
balancing is not concerned by cut ratio, selecting bad nodes instead
of random ones does not degrade intentionally the second metric. For
each node $v$ in a partition $i$, it is formally defined as:
\begin{equation}
  badness(v) = \frac{|\Gamma(v) \cap P(i))|}{|\Gamma(v)|},
\label{eq:badness}
\end{equation}
(the lower the worse). However, while improvement towards cut selects
the same amount of nodes in each machine,  the load-balancing improvement
heuristic only selects nodes on the most loaded machines.

The selected nodes are then migrated to a different machine. Again, 
heuristic improving cut proceeds while ignoring actual load on the machines: it migrates each selected
node to the partition $j$ containing the most of its neighbors:
\begin{equation}
j = \argmax_{i \in [k]} (|P(i) \cap \Gamma(v)|).
\label{eq:imp-cut}
\end{equation}
In contrast, the load balancing heuristic takes each machine load into account when
selecting a new partition for each node: for this it relies on the \baseline
strategy \eqref{eq:baseline}. Note that we can apply \baseline, as when this
heuristic is executed, the system deals with already received edges and nodes for reconfiguration.

Afterwards, both heuristics proceed the same way. They send the node to the
selected partition, and update \p regarding the node's new location.








Given the state of the partitioning at a given time, improving on one
criterion means reconfiguring the system, by moving a small part of
system nodes from one partition to another one.  The migration rate
must be small, in order not to impact the application running on top
of the partitioning.
The optimizer measures the application performance at runtime, decides when to
reconfigure, chooses a reconfiguration strategy, and commits it if it considers
this reconfiguration was successful. In order not to force the developers to
over instrument the application, and for overall simplicity/reusability, we
built a system that self-tunes solely based on one information given by the
application at runtime. This information takes the form of the average computation
time. From this value, the optimizer decides when to trigger a reconfiguration,
and measures after a reconfiguration if the average runtime is lower or not than
before a reconfiguration.


The optimization problem is then to reach a configuration, \ie a
certain graph partitioning, that minimizes request execution time at
the application level. As finding a particular graph partitioning is
NP-complete (\eg bisecting static graphs~\cite{NP}), we have to rely
on local search optimization. Considering our computing time feedback,
and two improvement criterions, a natural optimization framework is
\textbf{hill climbing}.  The difference of our setup with canonical
hill climbing is that we cannot instantly evaluate both neighbors of
current configuration, \ie the new configuration after a step on load
balancing and after a step on cut ratio. We thus make a random
choice towards one criterion, and act as a function of resulting
computing time.  We call this variant \textbf{blind hill climbing}.

\begin{figure}[t!]
\center
\includegraphics[width=0.75\linewidth]{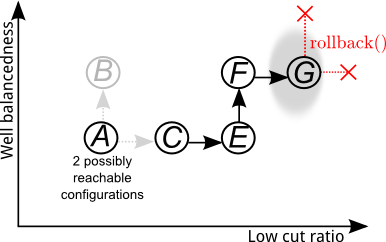}
\caption{Blind hill climbing optimization over current
  partitioning. \greedy leaves the system in configuration A. Ideal
  configuration is top-right (shaded area). From A, running a
  heuristic for improvement would lead to configuration B or C. Random
  coin flip selects heuristic to improve on cut ratio. Process is
  iterated (going from C to E, F and G), until none of the two
  heuristics can improve upon G. Rollbacks are operated for return to
  G, waiting for substantial graph growth before new optimization
  trials.}

        	\label{hill}
\end{figure}

This approach is summed-up by Figure~\ref{hill}; periodic optimizations
are conduced under trial and error. When no progress is possible in
any of the two directions, the configuration has reached a sweet spot
for the application performances.
Note that as nodes and edges arrive continuously at the datacenter,
this spot moves after each addition. Each optimization step then
performs in best effort fashion voluntarily considering current
graph as static.

\begin{figure*}[t]
     \begin{center}
       \subfigure[PL1000 graph, $k=4$.]{%
\hspace{-1.3cm}
	\includegraphics[width=0.37\linewidth]{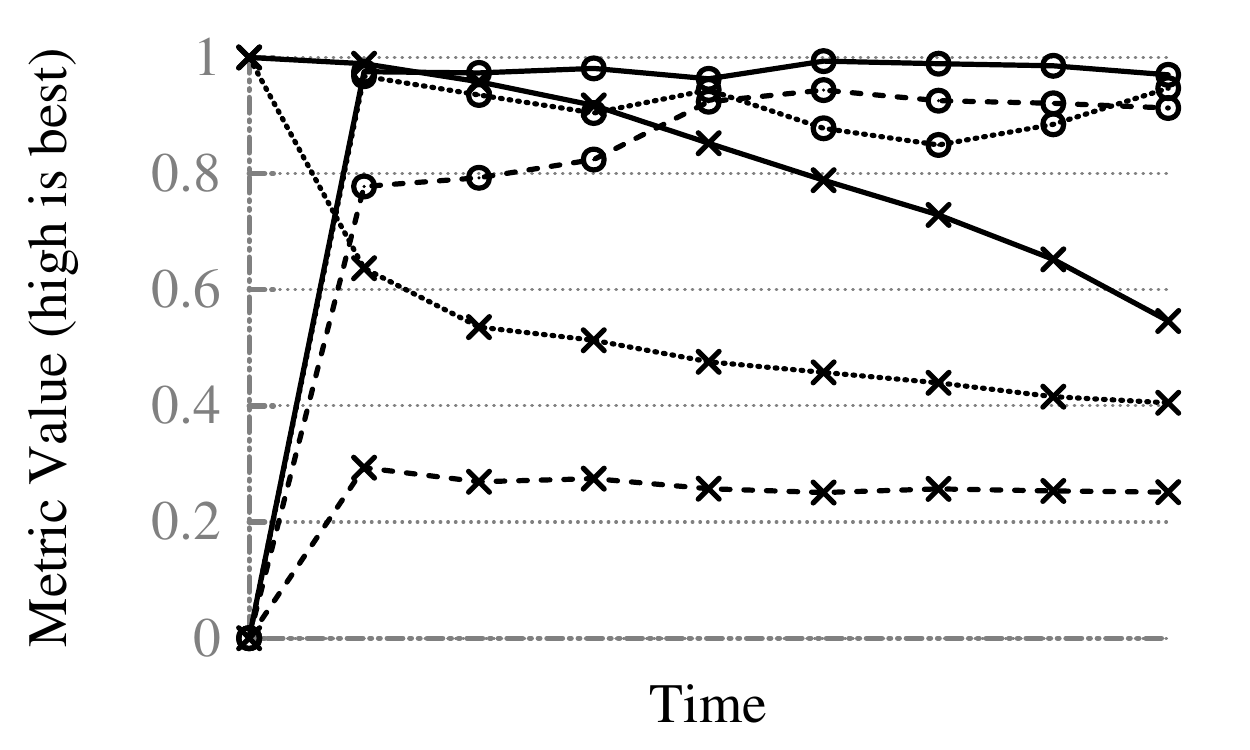}%
        \label{fig:pl}
        }%
        \subfigure[marvel graph, $k=8$.]{%
\hspace{-0.8cm}
	\includegraphics[width=0.37\linewidth]{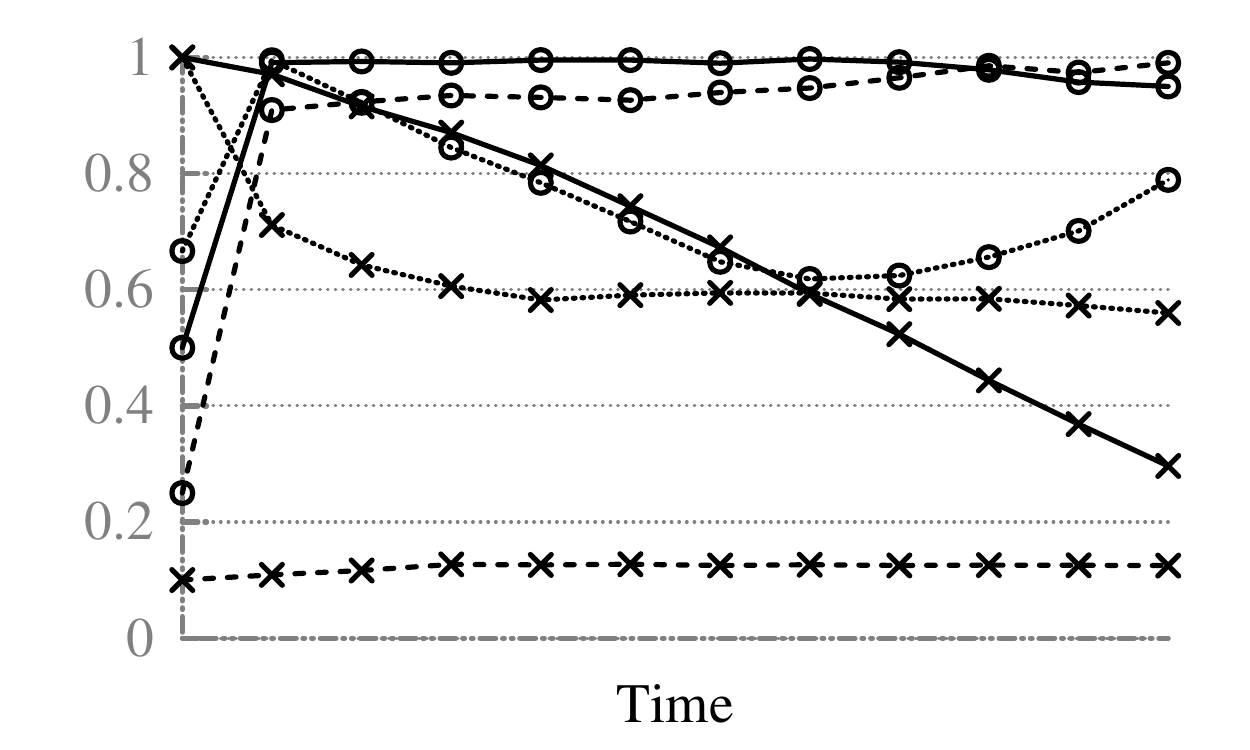}%
        \label{fig:marvel}
        }
        \subfigure[4elt graph, $k=4$.]{%
\hspace{-0.8cm}
	\includegraphics[width=0.37\linewidth]{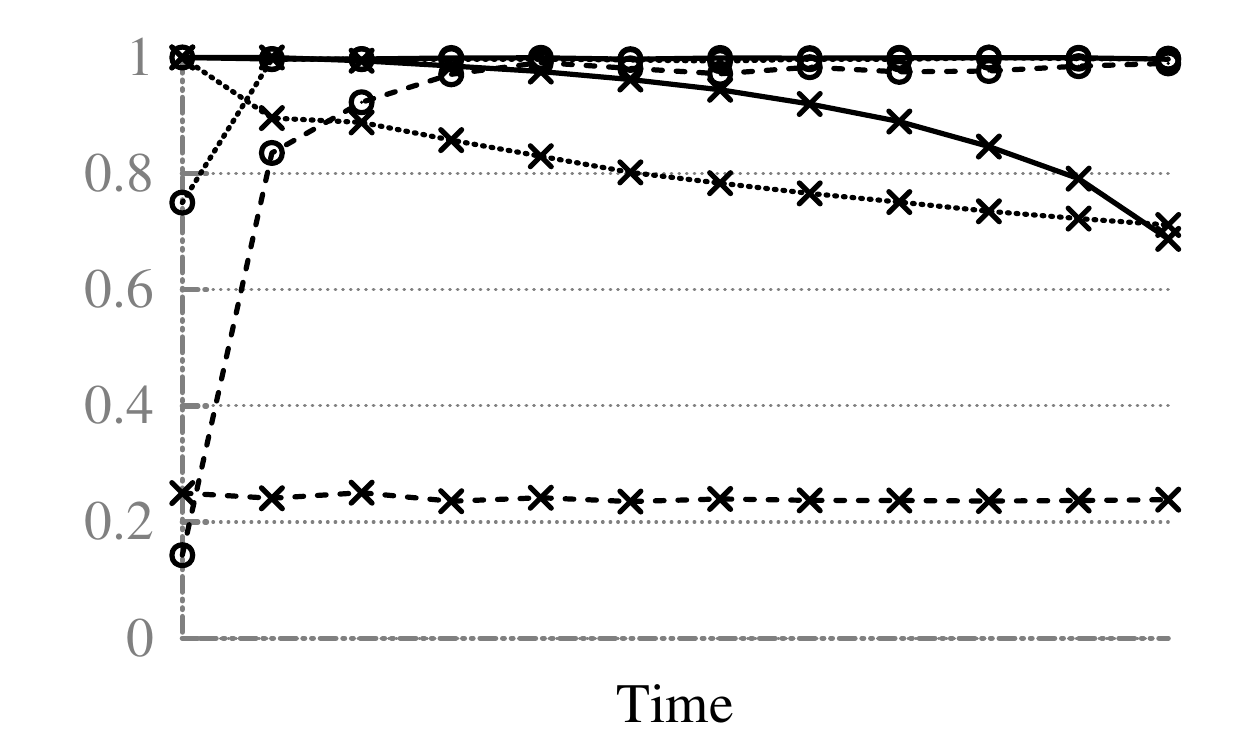}%
        \label{fig:4elt}
        }%
    \end{center}
    \caption{%
      \greedy partitioner (solid line), random (dashed line) and
      \baseline (dotted line) competitors, on 3 representative
      graphs. Load balancing (circles) and cut ratio (crosses) are
      plotted (the higher the better on the y-axis) for each approach,
      as the graph is streamed from its first edge to its last one
      (x-axis). }%
   \label{fig:greedy}
\end{figure*}

\begin{figure*}[t]
     \begin{center}
        \subfigure[PL1000]{%
          \hspace{-0.8cm}
	\includegraphics[width=0.4\linewidth]{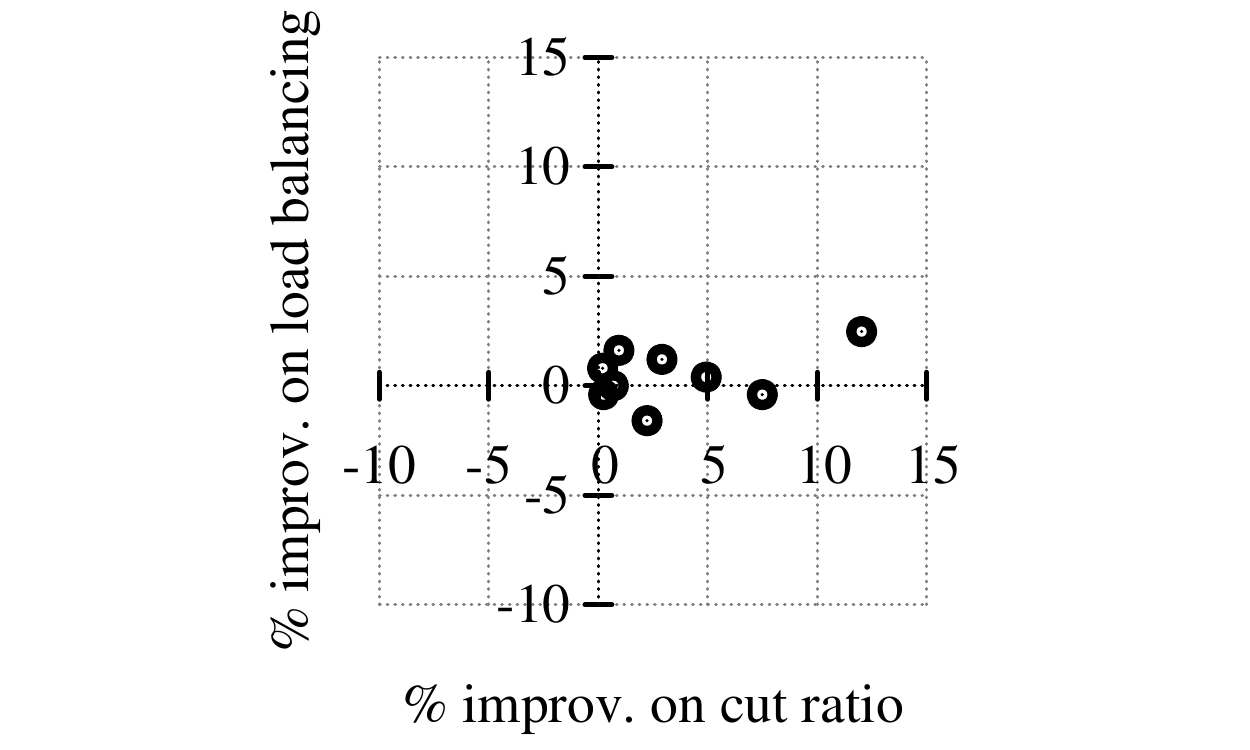}%
        \label{fig:imp-cut-pl}
        }%
        \subfigure[marvel]{%
          \hspace{-1.7cm}
	\includegraphics[width=0.4\linewidth]{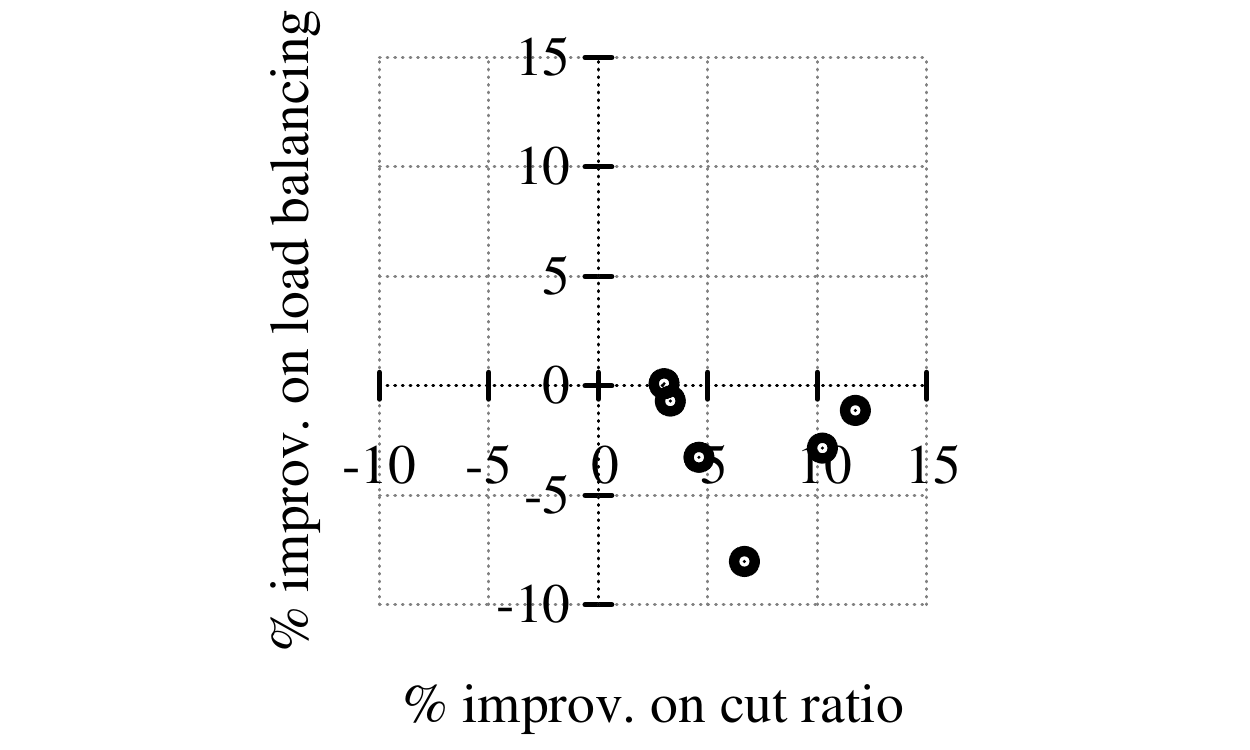}%
        \label{fig:imp-cut-marvel}
        }
        \subfigure[4elt]{%
          \hspace{-2cm}
	\includegraphics[width=0.4\linewidth]{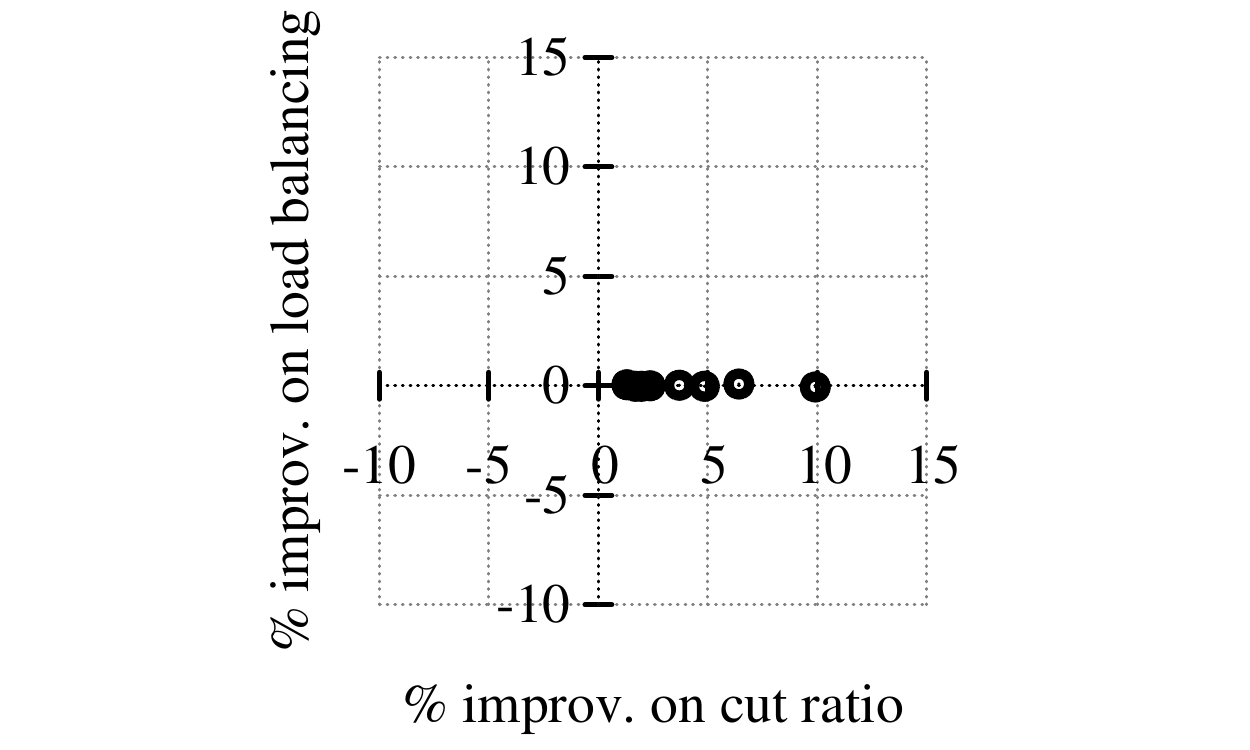}%
        \label{fig:imp-cut-4elt}
        }%
    \end{center}
    \caption{%
      Improvement on cut ratio and load balancing on greedily partitioned graphs, produced by successive runs seeking to improve \textbf{cut ratio}. Each point represents the percentage of improvement obtained considering current configuration and then executing heuristic to improve on cut. Ideally, points should follow the positive y-axis. The rightmost point on (a) means that a single call to heuristic improved previous cut by $12\%$ and also improving load balancing $2.6\%$ as a side effect.
     }%
   \label{fig:imp-cut}
\end{figure*}

\begin{figure*}[t!]
     \begin{center}
        \subfigure[PL1000]{%
          \hspace{-0.8cm}
	\includegraphics[width=0.4\linewidth]{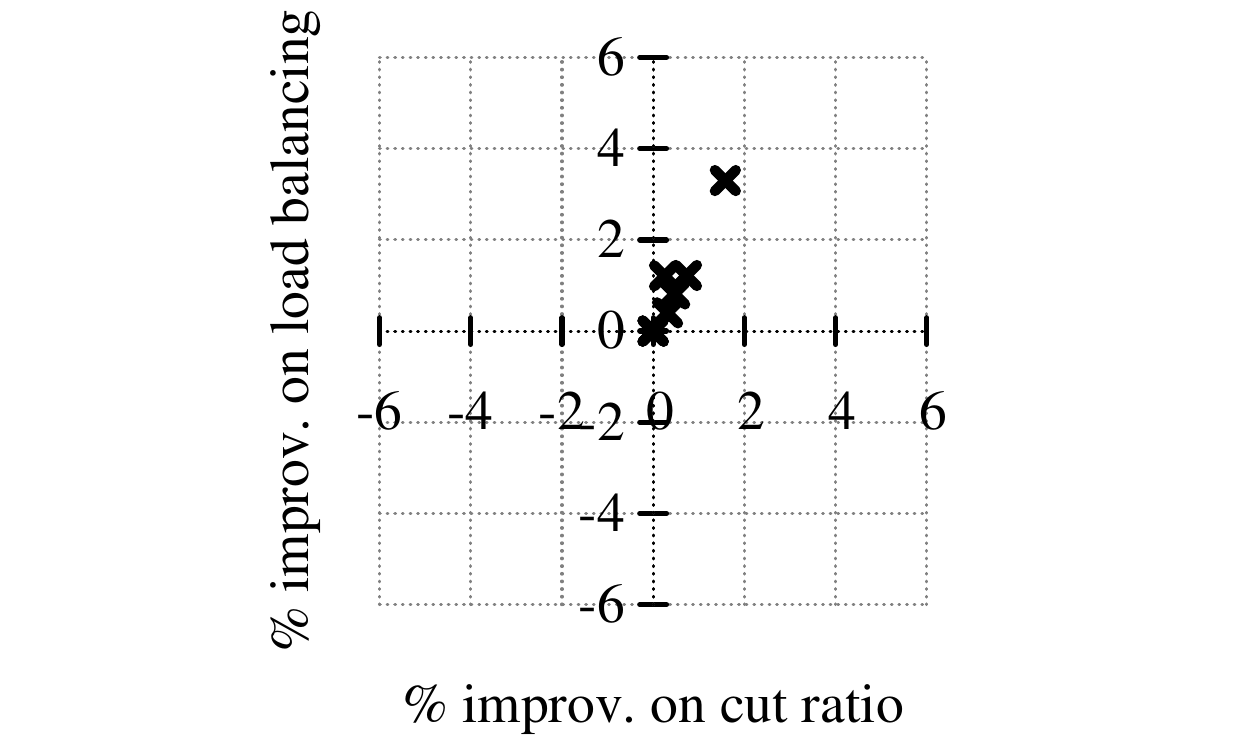}%
        \label{fig:imp-load-pl}
        }%
        \subfigure[marvel]{%
          \hspace{-1.7cm}
	\includegraphics[width=0.4\linewidth]{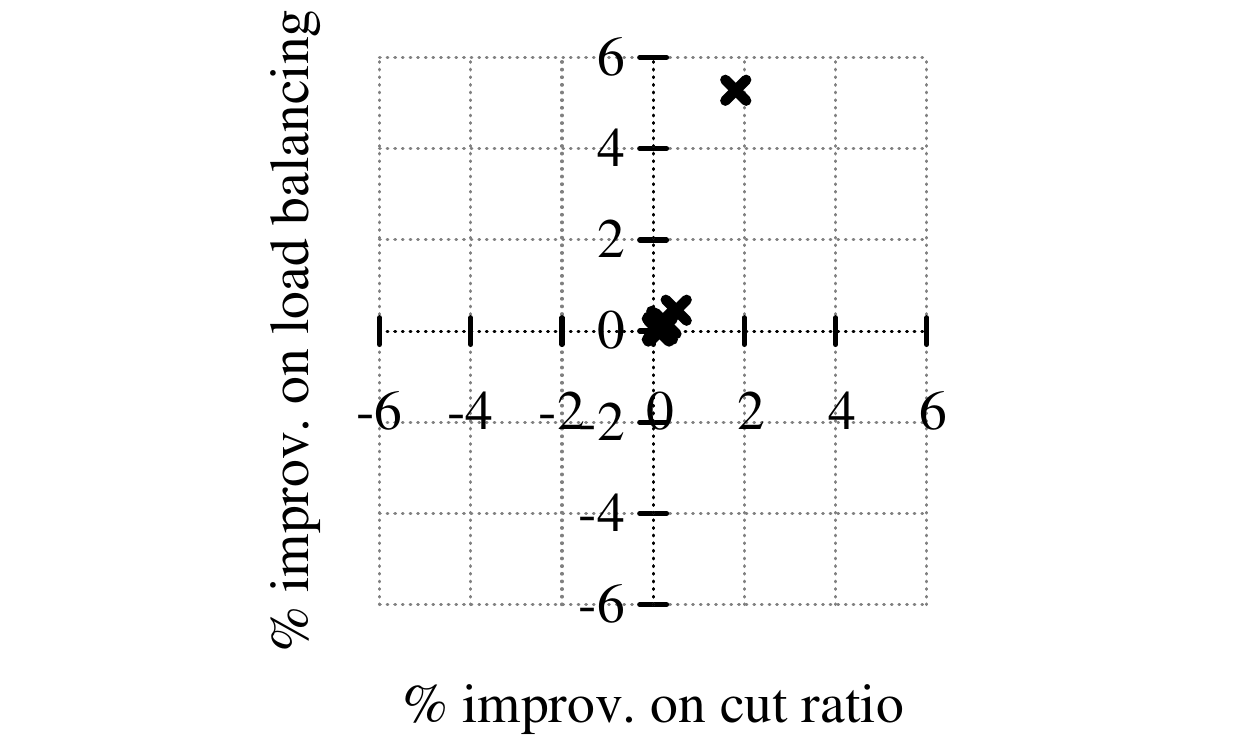}%
        \label{fig:imp-load-marvel}
        }
        \subfigure[4elt]{%
          \hspace{-2cm}
	\includegraphics[width=0.4\linewidth]{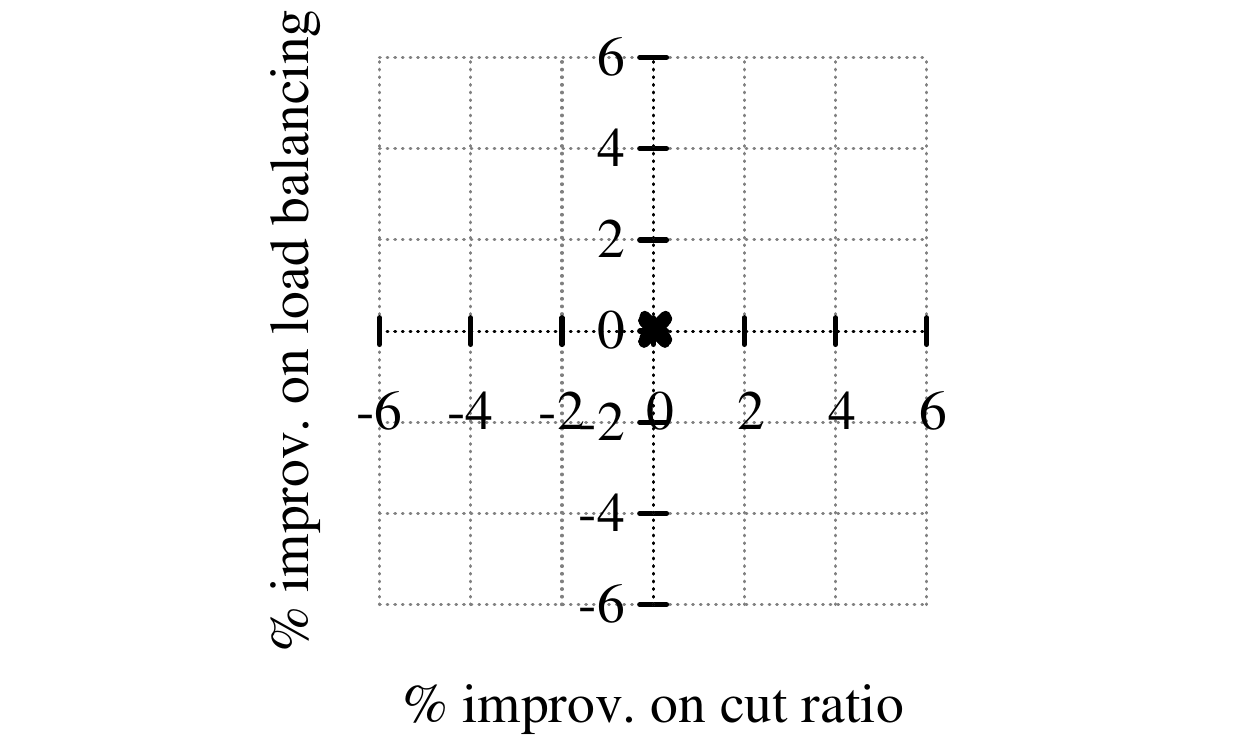}%
        \label{fig:imp-load-4elt}
        }%
    \end{center}
    \caption{%
      Improvement on min cut and load balancing, produced by successive runs seeking to improve \textbf{load balancing} Ideally, points should follow the positive x-axis (setting similar to Figure~\ref{fig:imp-cut}).
     }%
   \label{fig:imp-load}
\end{figure*}

Finally, note that we presented the optimizer as an independent
centralized entity for the sake of clarity. In practice, the optimizer
task is distributed on the set of machines hosting partitions.

\subsection{Simulation Results}
\label{sec:simulations}

\subsubsection{Simulation Setup}

For the sake of comparison with simulation results conduced
in~\cite{kdd12}, we use the same three graphs depicted as
representative from the major three types of network structures. The
first one, PL1000, is a synthetic graph of $1,000$ nodes with
clustered power-law characteristics\footnote{generated with NetworkX
  (http://networkx.github.io/)}. The social network, Marvel, consists
in $6,486$ characters, having $427,018$ interactions in comics. The
last graph, 4elt, of $15,606$ nodes and $45,878$ edges, is a FEM graph
from the NASA.

For all three datasets, as the graph is streamed in a random order
from the first edge to the last one. The number of partitions is
set to $4$ for PL1000, $8$ for Marvel and $4$ for 4elt, as
in~\cite{kdd12} for matters of reproducibility. Our greedy heuristic,
as well as random and \baseline results are exposed. Please note that
as \baseline assumes that when a node arrives at the partitioner, all
its (future) neighbor nodes are also known, we implement such
assumption in our simulation. This clearly gives it an advantage over
the two other approaches, but also serves as an indicator of the gap
between our greedy method and the one with full knowledge.

\subsubsection{stream-greedy performance}
\label{sec:greedy-part-perf}

We now assess the performances of random, \baseline and \greedy
heuristics by simulation, in a streamed graph setup.

Figure~\ref{fig:greedy} present results for well balancedness (lines
with rounds) and cut (lines with crosses) metrics on the y-axis, the
higher the better. The x-axis represents time, from $t=0$ where first
edge is dispatched, to $t_{end}$ where the whole graph has been
streamed (then having $G_{\infty}$ partitioned).

Regarding load balancing curves, results for all three graphs and all
three competitors are consistent and close to perfect balancing at
$t_{end}$, and all three are comparable.  Now regarding cut, we first
remark that \baseline results are consistent with results produced
in~\cite{kdd12} (where only cut ratio is considered), as operating
under the same assumptions. Produced results for random partitioning
are also consistent with theory, as random placement is awaited to cut
a fraction of $1-1/k$ of edges, thus $3/4$ for $k=4$ and $7/8$ for $k=8$.
The first learning is that baseline always beats random partitioning
(as seen in~\cite{kdd12}); this is also the case for our greedy
approach. Surprisingly, despite the fact that \baseline beats \greedy
on Marvel and slightly on 4elt graphs, \greedy outperforms the
deterministic greedy approach on PL1000. This shows that even with an
increased amount of information, \baseline does not perform better, as
it could be beaten by \greedy, even operating on less information. We
thus learn that operating under reduced assumptions still make
possible a very competitive first partitioning step, using an
intuitive partitioner such as \greedy.

We have presented and compared a simple one pass greedy stream
partitioner; as a consequence, edges placed by that heuristic are
never moved from one partition to another one. Next section shows
that we can periodically improve on one criterion or the other at
runtime.

\subsubsection{Improving on Cut Ratio}
\label{ss:cut}

Figure~\ref{fig:imp-cut} presents improvement percentages for the cut
criterion. Ideally, running a heuristic on one criterion (\eg on cut
ratio) should of course improve it, but also avoid degrading the
second one (\eg not degrade load balancing).  Simulations show that,
on the same three graphs at $t_{end}$, there is room for
improvement. With a typical value for top-$K$ worst nodes of $10\%$,
up to $12\%$ improvement is achieved at each procedure call, after
what next calls produce more slight changes. We also see that except
for few percent on the Marvel graph, improving on cut ratio does not
degrade current load balancing of graphs.

\subsubsection{Improving on Load Balancing}

Figure~\ref{fig:imp-load} presents results. Improvement on load
balancing at each call is positive but less important that for improvement
on the cut ratio criterion. This is explained by the fact that at
$t_{end}$, as seen on Figure~\ref{fig:greedy}, load balancing is
already close to perfect.
Those calls does not degrade the other criterion either.


\section{Illustration: graph-based instant recommender}
\label{illustration}

We now present a direct application of this optimization framework in
a system leveraging streamed graphs, detail implementation needs, and
show performance indicators.

\subsection{System Implementation of Blind Hill Climbing}

Regardless of the application relying on the partitioning process, few
system primitives are required to implement the optimization
framework. Here is a list:

\begin{algorithm}[t!]
    \While{True}{
    $c_{before} \leftarrow$ \texttt{getComputeTime()}\;
    \texttt{snapshot()}\;
    \texttt{buffer()}\;
        \eIf {\texttt{Random(cut,balancing) == cut}} { \texttt{OptimizeOnCut()} } { \texttt{OptimizeOnBalancing()} }
        $c_{after} \leftarrow$ \texttt{getComputeTime()}\;
        \eIf {$c_{after} > c_{before} + \epsilon * c_{before}$} {
          \texttt{rollback()}\;
        } {
          \texttt{commit()}\;
        }
        \texttt{flushBuffer()};\
  }
\vspace{0.2cm}
\caption{\footnotesize Blind hill climbing optimization, a generic approach for
  improving computation time in stream-enabled applications.}
\label{hill-alg}
\end{algorithm}

\begin{itemize}
\item \texttt{snapshot()}: atomically records node/edge repartition over machines.
\item \texttt{getComputeTime()}: top application returns current average compute time.
\item \texttt{commit()}: remain in current configuration, and free structure used for snapshoting.
\item \texttt{rollback()}: return to previous configuration (snapshoted earlier).
\item \texttt{buffer()}: record all incoming events (edges, requests) in a message queue. \texttt{flushBuffer()} consumes those buffered events.
\end{itemize}
From those primitives, we propose the following heuristic, that pursue
a hill climbing optimization based on application feedback (see
Algorithm~\ref{hill-alg}).

As configuration switch has practical costs, system parameter
$\epsilon$ denotes the degradation threshold over which it is
profitable to rollback to previous configuration. This threshold also
masks the slight deviation in average computing time due to particular
request patterns, or due to the arrival of nodes and edges in between
two optimizations.

\begin{figure*}[t!]
     \begin{center}
        \subfigure[Actual improvement of blind hill climbing optimization over a real dataset (MovieLens, $k=8$).]{%
\includegraphics[width=0.43\linewidth]{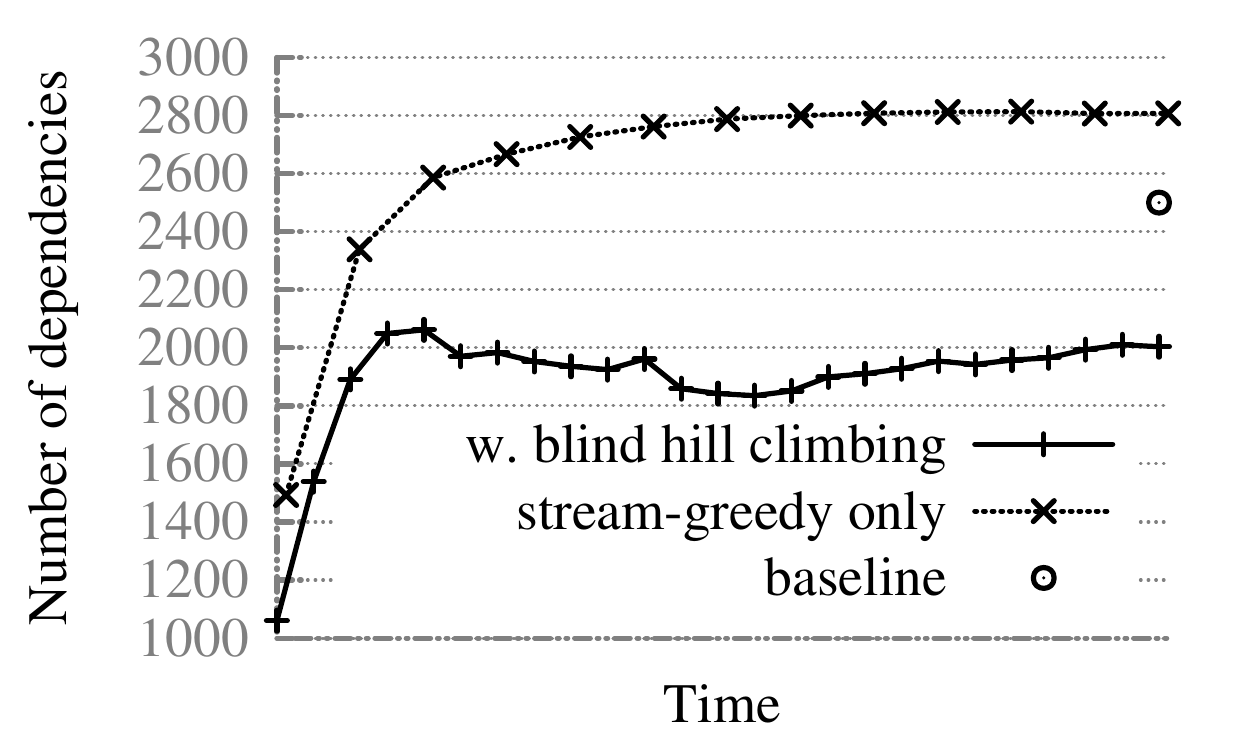}
        	\label{fig:sim-hill}
        }%
\hspace{1cm}
        \subfigure[Success of blind hill climbing optimization: each point on the two outer lines are successes, while inner lines represent failed optimization steps.]{%
\includegraphics[width=0.43\linewidth]{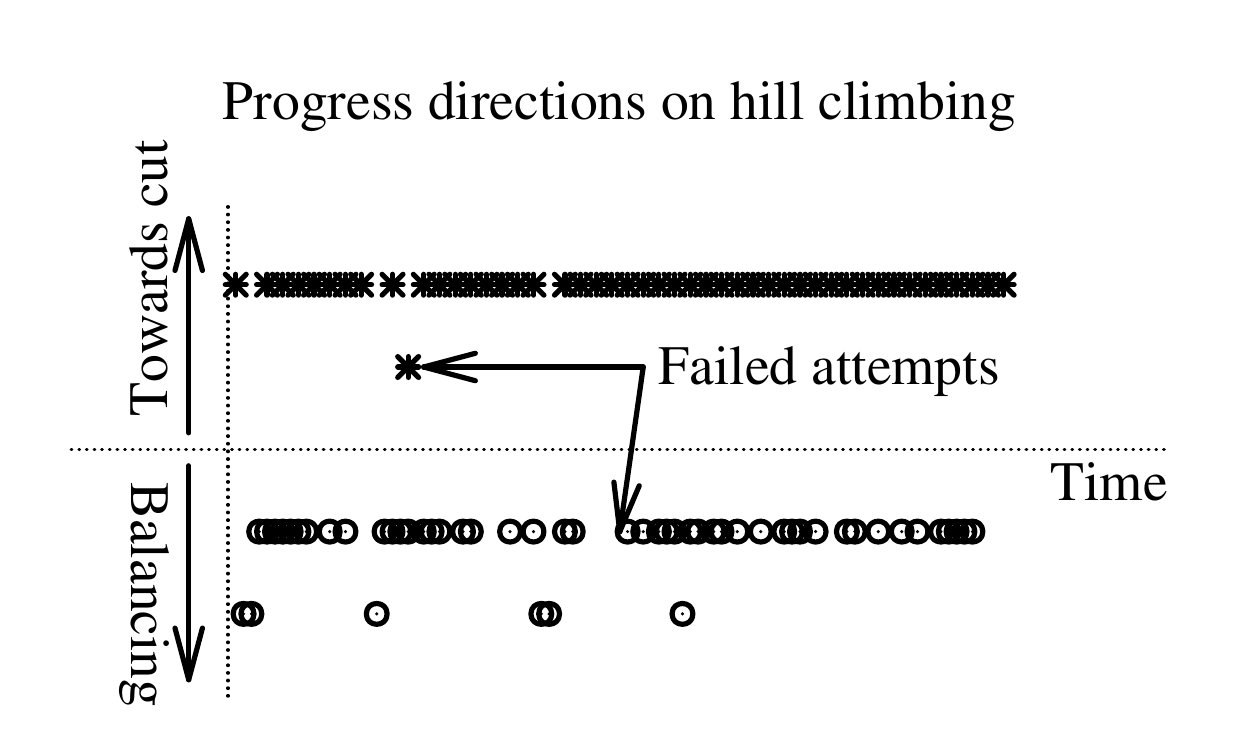}
        	\label{fig:sim-hill-success}
        }
      \end{center}
\caption{Simulation of the blind hill climbing optimization framework, over the streamed MovieLens dataset.}
   \label{fig:hill-expes}
\end{figure*}

\subsection{A System for Instant Recommendation}

A typical example of a latency critical service is recommendation.
The reactivity capabilities of the application to process fresh data
is key for successful services, as accurate and instant recommendation
based on previous clicks for instance~\cite{amazon}.  We take the
scenario of movie recommendation for the rest of this paper. In this
framework, nodes are users/movies and an edge is present if one user
has rated a particular movie. We are interested in treating ratings as
a stream, as their direct incorporation into the system for instant
recommendation can for instance solve the cold start problem at user
registration (\ie before waiting for the platform to compute offline
for later personalized recommendations).

Bahmani et al.~\cite{perso-pagerank} show that a variant of Pagerank,
based on random walks, provides fast and personalized
recommendations. On a user/item graph, it essentially consists in
launching multiple (damping) random walks from the node (user) that
needs a recommendation; most visited nodes (items) are extracted
through higher Pagerank values and sorted. On such a graph, top-nodes
(items) are good recommendations for that user.

As random walks are locality critical in the context of parallel computing,
we expect partitioning to be crucial for the recommender
performances. With a good partitioning (high density of links within
machines and low edge cut), launched damping random walks stay on the
same machine\footnote{We prototyped a Storm-based system~\cite{storm}
  implementing this recommendation engine, and answering to PUT/GET
  queries on the MovieLens graph. While adding and edge is made
  instantly, average computing time for single sequential GET
  recommendations is around $150$ms (using as much as $50,000$ random
  walks per recommendation), over a Xeon E5-2603 CPU (DELL
  T5600). This motivates the will to keep computation local to a
  machine for both speed and capability to handle many concurrent
  requests.}. If not, then many dependencies arise from inter-machines
communications.

\subsection{Simulation Over the MovieLens Dataset}

We ran personalized Pagerank for recommendation over the MovieLens
dataset~\cite{movielens}, a user/movie graph consisting of $100,000$
ratings from $1,000$ users on $1,700$ movies.

The blind hill climbing optimization, described on
Algorithm~\ref{hill-alg}, is implemented by a logical optimizer in the following way. \p
controls the frequency of optimizations. When \p decides to perform an
optimization step, it flips a coin to choose on which criterion to try
to improve. It then stops consuming events by calling the
\texttt{buffer()} primitive (architecturally this correspond in
practice to let the message queue store events,
\eg~\cite{kestrel}, and let it grow for an instant). Order to call improvement
heuristic is broadcasted to the $k$ machines.  Bad nodes are computed
(according to metric \eqref{eq:badness}) on each machine, in order not
to add computational burden onto the partitioner. On this simulation
$10\%$ of the worst nodes of each partition are migrated in each
optimization cycle.  Once optimization has completed on all machines,
a generic batch of requests is artificially executed on the system. If
\texttt{getComputeTime()} returns a degraded average completion time,
the partitioner broadcasts a rollback order (a commit one
otherwise). The rollback to previous partitioning configuration is
executed in a lockstep at each machine, by a simple memory overwrite
from previous \texttt{snapshot()} result. \p then returns in service
mode by calling \texttt{flushBuffer()}. This simple implementation
constitutes a coarse grain handling of system states; we leave a finer
grain handling of consistency for future work.

The primary metric to assess performance of the application related to
a given partitioning is the number of dependencies. A dependency of
the application occurs when a random walk (launched at any machine)
has to get onto another machine to pursue process. It is a direct
indicator of the latency at the application level, as network costs
clearly overcome local CPU computation.

Each simulated request triggers $1,000$ random walks (with a min-hop
of $3$ away from the querying user, and a damping factor of
$\alpha=0.9$). To simulate the increase of application-usage as the
graph grows, we ran a number of user requests of $1\%$ times the
current graph size, after each increase of the number of edges by
$5\%$. With an optimization step being triggered when graph size
increases by $5\%$, $\epsilon=1\%$, and $k=4$, dependencies are plotted on
Figure~\ref{fig:sim-hill}, as they evolve while the graph is streamed.
The hill climbing optimization clearly maintains dependencies at a
lower level than \greedy (close to $30\%$ less). The static
partitioning resulting from \baseline performs again slightly better
than \greedy, but as its operation is not periodically optimized, it
cannot compete with our blind hill climbing approach.

We now plot the on Figure~\ref{fig:sim-hill-success} the
success/failure of calls to improvement heuristics. For $100$
optimizations on a random criterion, $93$ are successful on cut ratio
criterion, and $6$ on load balancing. There is only $1$ rollback on
cut, but $47$ on load balancing. The more important failure rate over
load balancing is due to the already very good balance achieved
without optimization, as seen on Figure~\ref{fig:greedy} for other
graphs with \greedy. A solution to decrease this failure rate is to
bias the random choice towards the more successful of the two
criterions (\ie learn and call the most successful one with a higher
probability).

In conclusion, there is a clear advantage in periodically
reconfiguring current partitioning by calling lightweight optimization
procedures, for correcting past greedy decisions in the context of
streamed graphs. The framework we propose only takes as input the
application feedback and allows self-tuning in an efficient way.




\section{Related Work}
\label{related}

Static partitioning approaches take a graph as input and propose a
bisection as output (known as the minimum bisection problem).
Reaching a configuration with minimal number of inter-machine edges
while balancing the output is well known to be NP-complete~\cite{NP}.
This as been extended to $k$-partitioning, where the input graph is partitioned into
$k$ pieces~\cite{balanced}.
Methods for partitioning largely depend on the application using the
produced partitions, as computing while partitions fit network
architecture \cite{ipdps_part}, minimizing interactions between
storage servers~\cite{engines}, or computing over embarrassingly
parallel datasets~\cite{pregel} for instance.

Approaches like GraphChi~\cite{graphchi} or
TurboGraph~\cite{turbograph} aim at computing metrics over large
graphs on a centralized setting; they differ from stream-based approaches as they compute offline over the dataset and do not consider graph
updates for low latency operation and online request handling.

Staton et al.~\cite{kdd12} are the first to consider a stream of nodes
to be placed onto partitions on the fly. Many heuristics are proposed
and tested, from intuitive ones (considering balance) to more advanced
ones (considering clustering coefficient); we re-implement the best
performing one in this paper to compare it to our proposal. All
approaches are one pass; a placed vertex is never moved
afterward. Their paper assumes a full knowledge model, where the graph
to be streamed has to be present on one machine prior to the execution
of the proposed heuristics.

\section{Conclusion}
\label{conclusion}

This paper has exposed the hardness of partitioning a streamed graph
not already present on a computing device. A greedy partitioner taking
as input stream of edges has been proposed, that can compete with a
state of the art heuristic for partitioning under full
knowledge. While its operation is satisfying, we show that it is of
interest to periodically call an optimization procedure to improve upon
current partitioning, on the edge cut ratio or on the load balancing
criterions. These building blocks form a general optimization framework
allowing for application self-tuning, based solely on feedback on
computing time. An interesting question for future work is to formally
ask if there exists a greedy algorithm having provable bounds for its
efficiency to partition a streamed graph, in the classic stream
processing model.

\bibliographystyle{IEEEtran}
\bibliography{biblio}

\begin{thebibliography}{10}
\providecommand{\url}[1]{#1}
\csname url@samestyle\endcsname
\providecommand{\newblock}{\relax}
\providecommand{\bibinfo}[2]{#2}
\providecommand{\BIBentrySTDinterwordspacing}{\spaceskip=0pt\relax}
\providecommand{\BIBentryALTinterwordstretchfactor}{4}
\providecommand{\BIBentryALTinterwordspacing}{\spaceskip=\fontdimen2\font plus
\BIBentryALTinterwordstretchfactor\fontdimen3\font minus
  \fontdimen4\font\relax}
\providecommand{\BIBforeignlanguage}[2]{{%
\expandafter\ifx\csname l@#1\endcsname\relax
\typeout{** WARNING: IEEEtran.bst: No hyphenation pattern has been}%
\typeout{** loaded for the language `#1'. Using the pattern for}%
\typeout{** the default language instead.}%
\else
\language=\csname l@#1\endcsname
\fi
#2}}
\providecommand{\BIBdecl}{\relax}
\BIBdecl

\bibitem{engines}
\BIBentryALTinterwordspacing
J.~M. Pujol, V.~Erramilli, G.~Siganos, X.~Yang, N.~Laoutaris, P.~Chhabra, and
  P.~Rodriguez, ``The little engine(s) that could: scaling online social
  networks,'' \emph{IEEE/ACM Trans. Netw.}, vol.~20, no.~4, pp. 1162--1175,
  Aug. 2012. [Online]. Available:
  \url{http://dx.doi.org/10.1109/TNET.2012.2188815}
\BIBentrySTDinterwordspacing

\bibitem{perso-pagerank}
B.~Bahmani, A.~Chowdhury, and A.~Goel, ``Fast incremental and personalized
  pagerank,'' in \emph{VLDB}, 2010.

\bibitem{kdd12}
I.~Stanton and G.~Kliot, ``Streaming graph partitioning for large distributed
  graphs,'' in \emph{KDD}, 2012.

\bibitem{mr}
\BIBentryALTinterwordspacing
J.~Dean and S.~Ghemawat, ``Mapreduce: simplified data processing on large
  clusters,'' \emph{Commun. ACM}, vol.~51, no.~1, pp. 107--113, Jan. 2008.
  [Online]. Available: \url{http://doi.acm.org/10.1145/1327452.1327492}
\BIBentrySTDinterwordspacing

\bibitem{storm}
N.~Marz, ``{Storm Project},'' \url{http://storm-project.net/}, 2012.

\bibitem{milky}
T.~Akidau, A.~Balikov, K.~Bekiroglu, S.~Chernyak, J.~Haberman, R.~Lax,
  S.~McVeety, D.~Mills, P.~Nordstrom, and S.~Whittle, ``Millwheel:
  Fault-tolerant stream processing at internet scale,'' in \emph{VLDB}, 2013,
  pp. 734--746.

\bibitem{kinesis}
Amazon, ``{Amazon Kinesis},'' \url{http://aws.amazon.com/kinesis/}, 2013.

\bibitem{balanced}
K.~Andreev and H.~R\"{a}cke, ``Balanced graph partitioning,'' in \emph{SPAA},
  2004.

\bibitem{pregel}
G.~Malewicz, M.~H. Austern, A.~J. Bik, J.~C. Dehnert, I.~Horn, N.~Leiser, and
  G.~Czajkowski, ``Pregel: a system for large-scale graph processing,'' in
  \emph{SIGMOD}, 2010.

\bibitem{NP}
M.~R. Garey and D.~S. Johnson, \emph{Computers and Intractability; A Guide to
  the Theory of NP-Completeness}.\hskip 1em plus 0.5em minus 0.4em\relax New
  York, NY, USA: W. H. Freeman \& Co., 1990.

\bibitem{amazon}
G.~Linden, B.~Smith, and J.~York, ``Amazon.com recommendations: Item-to-item
  collaborative filtering,'' \emph{IEEE Internet Computing}, vol.~7, no.~1, pp.
  76--80, Jan. 2003.

\bibitem{movielens}
J.~L. Herlocker, J.~A. Konstan, A.~Borchers, and J.~Riedl, ``An algorithmic
  framework for performing collaborative filtering,'' in \emph{SIGIR}, 1999.

\bibitem{kestrel}
B.~Cook, ``{Kestrel distributed message queue},''
  \url{https://github.com/robey/kestrel/}, 2013.

\bibitem{ipdps_part}
D.~Ajwani, S.~Ali, and J.~Morrison, ``Graph partitioning for reconfigurable
  topology,'' in \emph{IPDPS}, 2012.

\bibitem{graphchi}
A.~Kyrola, G.~Blelloch, and C.~Guestrin, ``Graphchi: large-scale graph
  computation on just a pc,'' in \emph{OSDI}, 2012.

\bibitem{turbograph}
W.-S. Han, S.~Lee, K.~Park, J.-H. Lee, M.-S. Kim, J.~Kim, and H.~Yu,
  ``Turbograph: a fast parallel graph engine handling billion-scale graphs in a
  single pc,'' in \emph{KDD}, 2013.

\end{thebibliography}

\end{document}